\documentclass[12pt]{article}

\usepackage{latexsym,amsmath,amssymb}
\usepackage{citesort}

\newcounter{subequation}[equation]
\makeatletter

\def\thesubequation{\theequation\@alph\c@subequation}
\def\@subeqnnum{{\rm (\thesubequation)}}
\def\slabel#1{\@bsphack\if@filesw {\let\thepage\relax
   \xdef\@gtempa{\write\@auxout{\string
      \newlabel{#1}{{\thesubequation}{\thepage}}}}}\@gtempa
   \if@nobreak \ifvmode\nobreak\fi\fi\fi\@esphack}
\def\subeqnarray{\stepcounter{equation}
\let\@currentlabel=\theequation\global\c@subequation\@ne
\global\@eqnswtrue
\global\@eqcnt\z@\tabskip\@centering\let\\=\@subeqncr
$$\halign to \displaywidth\bgroup\@eqnsel\hskip\@centering
  $\displaystyle\tabskip\z@{##}$&\global\@eqcnt\@ne
  \hskip 2\arraycolsep \hfil${##}$\hfil
  &\global\@eqcnt\tw@ \hskip 2\arraycolsep
  $\displaystyle\tabskip\z@{##}$\hfil
   \tabskip\@centering&\llap{##}\tabskip\z@\cr}
\def\endsubeqnarray{\@@subeqncr\egroup
                     $$\global\@ignoretrue}
\def\@subeqncr{{\ifnum0=`}\fi\@ifstar{\global\@eqpen\@M
    \@ysubeqncr}{\global\@eqpen\interdisplaylinepenalty \@ysubeqncr}}
\def\@ysubeqncr{\@ifnextchar [{\@xsubeqncr}{\@xsubeqncr[\z@]}}
\def\@xsubeqncr[#1]{\ifnum0=`{\fi}\@@subeqncr
   \noalign{\penalty\@eqpen\vskip\jot\vskip #1\relax}}
\def\@@subeqncr{\let\@tempa\relax
    \ifcase\@eqcnt \def\@tempa{& & &}\or \def\@tempa{& &}
      \else \def\@tempa{&}\fi
     \@tempa \if@eqnsw\@subeqnnum\refstepcounter{subequation}\fi
     \global\@eqnswtrue\global\@eqcnt\z@\cr}
\let\@ssubeqncr=\@subeqncr
\@namedef{subeqnarray*}{\def\@subeqncr{\nonumber\@ssubeqncr}\subeqnarray}
\@namedef{endsubeqnarray*}{\global\advance\c@equation\m@ne%
                           \nonumber\endsubeqnarray}

\renewcommand\maketitle{\par
  \begingroup
    \if@twocolumn
      \ifnum \col@number=\@ne
        \@maketitle
      \else
        \twocolumn[\@maketitle]%
      \fi
    \else
      \newpage
      \global\@topnum\z@   
      \@maketitle
    \fi
    \thispagestyle{plain}\@thanks
  \endgroup
  \setcounter{footnote}{0}%
  \global\let\thanks\relax
  \global\let\maketitle\relax
  \global\let\@maketitle\relax
  \global\let\@thanks\@empty
  \global\let\@author\@empty
  \global\let\@date\@empty
  \global\let\@title\@empty
  \global\let\title\relax
  \global\let\author\relax
  \global\let\date\relax
  \global\let\and\relax
}

\makeatother

\DeclareFontFamily{OT1}{rsfs11}{}
\DeclareFontShape{OT1}{rsfs11}{m}{n}{ <-> rsfs11 }{}
\DeclareMathAlphabet{\mathscript}{OT1}{rsfs11}{m}{n}

\newcommand{\gtlt}{\mathrel{\raise2.5pt\hbox{\oalign{$\scriptstyle>$\crcr
$\scriptstyle<$}}}}

\newcommand{\e}{{\mathrm e}}

\newcommand{\ie}{{\it i.e.}\ }

\newcommand{\pt}{\partial}

\newcommand{\be}{\begin{equation}}
\newcommand{\ee}{\end{equation}}
\newcommand{\nn}{\nonumber}
\newcommand{\bea}{\begin{eqnarray}}
\newcommand{\eea}{\end{eqnarray}}
\newcommand{\bsea}{\begin{subeqnarray}}
\newcommand{\esea}{\end{subeqnarray}}

\def\a{\alpha}

\def\g{\gamma}

\def\d{\delta}
\def\e{\epsilon}

\def\f{\phi}

\def\l{\lambda}
\def\m{\mu}
\def\n{\nu}

\def\r{\rho}
\def\s{\sigma}

\def\t{\tau}

\def\G{\Gamma}

\begin{document}

\begin{titlepage}

\begin{flushright}
Imperial/TP/041201 \\
KUL-TF-04/41\\
hep-th/0501212
\end{flushright}
\vspace{.5cm}
\begin{center}
\baselineskip=16pt
{\LARGE    Stability of Ho\v{r}ava-Witten Spacetimes }\\
\vspace{10mm}
{\large J.-L. Lehners$^{1}$, P. Smyth$^{1,2}$ and K.S.
Stelle$^{1}$}

\vspace{15mm}

{\small $^1$ The Blackett Laboratory, Imperial College London} \\
 Prince Consort Road, London SW7 2AZ, U.K.
\\ \vspace{6pt}
$^2$ Instituut voor Theoretische Fysica - Katholieke Universiteit
Leuven
\\
Celestijnenlaan 200D B--3001 Leuven, Belgium
\end{center}
\vspace{15mm} \abstract{Ho\v{r}ava-Witten spacetimes necessarily include two
branes of opposite tension. If these branes are BPS we are led to a puzzle: a
negative tension brane should be unstable as it can loose energy by expanding,
whereas a BPS brane should be stable as it resides at a minimum of the energy. We
provide a detailed analysis of the energy of such braneworld spacetimes in
$5$ dimensions. This allows us to show by a non-perturbative positive energy
theorem that Ho\v{r}ava-Witten spacetimes are stable, essentially because the
dynamics of the branes is entirely accounted for by the behaviour of the bulk
fields. We also  perform an ADM perturbative Hamiltonian analysis at quadratic
order in order to illustrate the stability properties more explicitly.}

\vspace{2mm} \vfill \hrule width 3.cm \vspace{2mm}{\footnotesize
\noindent E-mail: \texttt{jean-luc.lehners@imperial.ac.uk,
paul.smyth@imperial.ac.uk, k.stelle@imperial.ac.uk} }
\end{titlepage}

\section{Introduction}

Ho\v{r}ava-Witten (HW) theory links $11$-dimensional supergravity on the
orbifold $S^1/\mathbb{Z}_2$ with strongly coupled heterotic $E_8 \times E_8$
string theory \cite{HW}. It suggests that as one probes to higher energy, our
$4$-dimensional world first goes through an intermediate regime where the
orbifold dimension becomes visible, the universe thus appearing $5$-dimensional
with two boundary branes \cite{BD,LOSW}. One of these branes holds us while the
second one could hold other matter that would appear ``dark'' to us. Only at
energies of the order of the string scale would the universe look
$11$-dimensional.

The intermediate $5$-dimensional energy regime has led to a large
number of new cosmological models, coming under the general name
of ``braneworlds''. We shall refer to such dual braneworld spacetimes generically
as HW spacetimes.\footnote{Explicit models have been considered as
supergravity brane solutions in $D=5$ after dimensional reduction from
$D=11$ in Ref.\
\cite{LOSW} and in the Randall-Sundrum scenarios based on branes in AdS
space in Refs \cite{RS}. A review from a cosmological viewpoint is given in
Ref.\ \cite{Davis}.} One of the distinguishing features of such HW spacetimes
is that their topology is a line element times a non-compact space, with two
branes residing at the boundaries of the line element. Usually the line
element is realised as the orbifold $S^1/\mathbb{Z}_2$, which can thus also
be viewed in the ``upstairs'' picture as a circle with
$\mathbb{Z}_2$ identifications. Another is that one brane has
positive tension while the other has the opposite negative
tension, owing to the fact that the orbifold direction is compact.
By construction, these spacetimes are supersymmetric and would thus
seem to be stable. However, the negative tension brane would appear
to give rise to ``ballooning'' modes, \ie one would expect
that it could loose energy by expanding, thus becoming unstable.\footnote{Indeed,
at Stephen Hawking's $60^{\rm th}$ birthday conference, Brandon Carter famously
exhibited this mode in a talk warning that one should pay careful attention to
the possibility of instabilities in braneworld spacetimes, putting the lecturer's
pointer stick under compression, and inadvertently snapping it in two. This
paper is thus a reply to Brandon's concern.} In this paper we aim to resolve this issue by providing a comprehensive analysis of the energy and the fluctuations of braneworld solutions. So, the main question addressed in this paper is --- which way does it go? Are supersymmetric HW spacetimes stable or unstable? In the end, we will answer this question in the affirmative.

We are going to consider perturbations about background solutions
of the following type:
\bea
 ds_{5}^{2}&=& e^{2A(y)}\eta_{\mu
\nu}dx^{\mu}dx^{\nu}+e^{-8A(y)}dy^2\ ,\label{eq1}  \\
\phi &\sim& ln{H(y)}, \, \, H=k|y|+c\ ,\label{eq2}
\eea
where $\phi$ is
the scalar field supporting the branes and $H$ is a linear
harmonic function. The $\mathbb{Z}_2$ identification appears as a
symmetric kink in the harmonic function at the location of the
branes. Intuitively, there are two basic types of motion that the
branes can perform: a centre-of-mass motion of the set of two branes and a
relative motion of the branes with respect to each other. The latter is
described by the radion mode $r(x^{\mu})$ which we can write as
\bea
 ds_{5}^{2}&=& e^{r(x^{\r})+2A(y)}g^{(4)}_{\mu
\nu}dx^{\mu}dx^{\nu}+e^{-2r(x^{\r})-8A(y)}dy^2\ ,\label{eq3}  \\
\phi &\sim& ln{H(y)}+r(x^{\r})\ .\label{eq4}
\eea
One can consistently
truncate the theory to a system containing this mode coupled to
$4$-dimensional gravity \cite{LS,LLP}, the equations of motion then being
\bea
R^{(4)}_{\mu \nu}&=&c' \pt_{\mu}r \pt_{\nu}r  \\
\Box^{(4)}{r}&=&0
\eea
where $c'$ is a positive constant.
The kinetic term for the radion mode hence appears with a positive
sign, and this indicates that there should
be no instability associated with it.\\

Let us now turn our attention to the centre-of-mass mode.
Goldstone modes can be written down either by looking at
diffeomorphisms moving the system relative to a fixed coordinate frame
\cite{ACGNR}. If we perform a coordinate transformation shifting the location of
the wall by a constant parameter $s$, physically not much has happened
in the original static solution (\ref{eq1},\ref{eq2}). Brane
solutions always break translational symmetries and a constant
shift merely moves the brane relative to an arbitrary fixed
reference coordinate frame. However, one can then promote the
modulus $s$ to a Goldstone mode by giving it a dependence on the
worldvolume coordinates $s(x^{\m})$, thus allowing {\it relative}
shifts of different points of the brane. The metric ansatz is then
given by \be ds_{5}^2=e^{2A(y-s(x^{\r}))}g^{(4)}_{\mu \nu}
dx^{\mu}dx^{\nu}+e^{-8A(y-s(x^{\r}))}dy^2. \label{slinky} \ee We
also set \be H=k|y-s(x^{\r})|+c\ .\label{slinkyharmonic} \ee The
centre-of-mass mode can be thought of in terms of the
$\mathbb{Z}_2$ identification being made local as a function of
the worldvolume coordinates. This mode therefore describes a sort
of shear or warping of the HW end branes.\footnote{In an
$S^1/\mathbb{Z}_2$ interpretation of the HW line element section
of the spacetime, one might envision this as the torsional mode
for a relative twisting of the coils in a child's ``slinky'' toy.}
In many treatments of braneworlds this mode is not discussed,
under the assumption that the $\mathbb{Z}_2$ symmetry projects it
out. However, only a global $\mathbb{Z}_2$ symmetry can project it
out, and from the standpoint of the supergravity theory,
there does not seem to be a clear motivation for making such an
arbitrary truncation. Letting the $\mathbb{Z}_2$ become local as a
function of the worldvolume coordinates preserves the upstairs/downstairs
pictures of the HW line element while allowing for an intuitively
natural mode of the HW system.

The mode described in the ansatz (\ref{slinky}), based on the curved background
solution (\ref{eq1},\ref{eq2}), should be contrasted with a putative centre-of-mass
mode in the $11$-dimensional HW picture. In that case, the bulk spacetime is
flat and thus, because of its translation invariance, shifting the bulk solution by
a constant produces no Goldstone mode since translating the background makes no
change in the local supergravity field values. Another way to see this is to note
that the $\mathcal{N}=1,\ D=10$ supermultiplet on the end-brane worldvolume does not
contain any scalar degrees of freedom. Thus, the very basic HW background solution
in $D=11$ spacetime consisting of flat space bounded by two $D=10$ flat boundaries
does not have a natural Goldstone mode and thus agrees with the string-theory
picture of an orientifold, without a translational mode. By contrast, the HW
spacetime in $D=5$ obtained after reduction from $D=11$ on a
Calabi-Yau manifold to $5$ dimensions has a curved bulk solution and translational
scalar modes are in this case present for the bounding branes \cite{LOSW}.

Including the translational mode
(\ref{slinky},\ref{slinkyharmonic}) and the radion mode
(\ref{eq3},\ref{eq4}) allows for the two intuitively obvious
motions of a set of two branes bounding a one-dimensional line
element: they can either move together (translation) with local
dependence on the worldvolume (warping) or oppositely (radion),
causing the line element's length to locally expand or contract as a function of
the worldvolume coordinates. Both the translational and the radion modes involve a
stated dependence on the line element's $x^5=y$ coordinate, but only the
latter gives a Kaluza-Klein consistent ``braneworld'' reduction
\cite{LS,LLP}. The translational mode can be considered in
isolation in a low-energy approximation, but, unlike the radion
mode, it couples to higher 5-dimensional modes at the trilinear
and higher orders. Accordingly, while we will find it instructive
to consider the energy positivity properties of the translational
and radion fluctuations (\ref{slinky},\ref{slinkyharmonic}) and
(\ref{eq3},\ref{eq4}) at the bilinear level (relevant to
linearised fluctuation equations), understanding the full story
requires use of a Witten-Nester positivity approach to the full
$D=5$ theory with boundaries.

We emphasise that both of the fluctuation modes (\ref{eq3},\ref{eq4}) and
(\ref{slinky},\ref{slinkyharmonic}) can be identified with fluctuations in the
bulk geometry. This is an indication that the dynamics of the bulk by itself
can give an accurate account of the physics of the system. In fact the
$\mathbb{Z}_2$ symmetry of the orbifold implies that the Israel matching
conditions at the locations of the branes become sets of boundary conditions
on the bulk fields.\footnote{This point has been emphasised in Ref.\
\cite{Moss} in a reformulation of the delta-function contributions to the
$D=10+1$ HW action and supersymmetry transformations.} We will be able to exploit
this property to reduce the dynamical description of the branes to that of the
nearby bulk spacetime. This is the key to showing that the energy of HW
spacetimes is positive by deriving a Witten-Nester energy positivity theorem
for the bulk.

In the next section we will review the HW solutions that we use here as
representative braneworld solutions. Our basic arena for this discussion will
be the $D=5$ dimensional reduction of type IIB supergravity as detailed in
\cite{BDLPS}. Thus, the basic $D=5$ brane solutions we will be working with
are generalisations of RS solutions. We will present the full action of
this setup, including the brane sources. We then proceed in section $3$
to define the energy, and show that in static gauge it suffices to look at
the stability properties of the bulk alone. This enables us to
prove a Witten-Nester positive energy theorem in section $4$ and thus establish
the stability of the $5$-dimensional HW spacetimes. In section $5$ we use an
ADM Hamiltonian approach to give an explicit calculation of the energy at
quadratic order in fluctuations by way of illustration. We conclude with a
discussion of our results.

\section{A Supersymmetric RS Solution}

We will briefly review the supersymmetric RS solution discussed in
\cite{CLP,DLS}. In the RS models \cite{RS} the $5$-dimensional
space consists of $Ad{S}_{5}$ space. When trying to embed the RS
models in string theory, it is therefore necessary to look for
theories which admit $Ad{S}_{5}$ vacua upon compactification. Type
IIB supergravity is known to have an $Ad{S}_{5} \times S^{5}$
vacuum solution, and hence $5$-sphere reductions of this theory
suggest themselves as the key to the problem. In the dimensional
reduction one promotes the volume modulus of the $5$-sphere to a
dynamical field. This ``breathing mode'' gives rise to a potential
in the uncompactified $5$ dimensions. This gravity plus scalar
system supports a domain wall solution, which can be identified
with the (positive tension) brane in the RS2 model, after a
$\mathbb{Z}_{2}$ identification of the background space at the
location of the brane. In fact the theory in $5$ dimensions can be
truncated consistently to a scalar-gravity theory with the simple
Lagrangian \cite{BDLPS}:
\be 
\mathcal{L}_{5}=\sqrt{-g}\,[R-\frac{1}{2}(\pt
\phi)^{2}-V(\phi)], \label{Lagrangian} 
\ee 
with 
\be 
V(\phi)=8m^2
e^{8\alpha \phi}-R_{5} e^{\frac{16\alpha}{5}\phi}\ ,
\ee 
where
$m$ and $R_{5}$ are constants and $\alpha=\frac{\sqrt{15}}{12}.$
Note that $\phi$ is not related to the dilaton in $10$ dimensions,
but instead represents the volume of the $5$-sphere compactification.

The domain wall solution in this theory, which we shall take to be our basic
example of a HW brane, is given by
\bea
ds_5^2 &=& (b_1 H^{2/7}+b_2 H^{5/7})^{1/2} \eta_{\mu \nu} dx^{\mu} dx^{\nu}  + (b_1 H^{2/7}+ b_2 H^{5/7})^{-2} dy^2\ ,\\
\phi &=& -\frac{\sqrt{15}}{7}ln(H), \, \, H= k|y|+c ,
\label{Solution}
\eea
with $ b_1=\pm \frac {28m}{3k}$ , $ b_2=\pm\frac{14}{15k}\sqrt{5R_5}$.
Here $k$ denotes the tension and $y$ denotes the direction transverse to the
brane. A second brane of opposite tension is placed at
$y=\pi$, where a $\mathbb{Z}_2$ identification between $y=\pm\pi$ is made. Thus
the topology of the full spacetime is $\mathbb{R}_4 \times
S^1/\mathbb{Z}_2$. One has to choose $b_2>0$ and $b_1<0$ in order for there
to exist a $k\rightarrow 0$ pure AdS limit, thus obtaining an RS scenario
\cite{LLP,DLS}. Note that
$k$ positive (a ``trough'' harmonic function) corresponds to a
negative tension brane, as can be verified using the Israel
matching conditions. In order for the metric to be real, we need
\be
H(y)^{\frac{3}{7}}>\mid \frac{b_1}{b_2}\mid, \label{Reality}
\ee
and we therefore choose the integration constant $c$ accordingly.

\subsection*{Coupling to Brane Actions}

The domain wall solution (\ref{Solution}) yields the following
singular terms in the Einstein tensor and in the scalar field
equation (all non-singular terms are denoted $\it Reg$ and solve the
bulk field equations) \cite{CDLLPS}:
\bea
G_{\m \n} &=& \frac{3 k }{14}(2 b_1 H^{-\frac{5}{7}}+ 5 b_2
H^{-\frac{2}{7}})(g_{55})^{-\frac{1}{2}}g_{\m \n} [\d(y)-\d(\pi-y)] + {\it Reg}
\label{5Dsingular1} \\ G_{yy} &=& 0 + {\it Reg} \label{5Dsingular2} \\ \Box \f
&=& -\frac{2
\sqrt{15}k}{7}(b_1 H^{-\frac{5}{7}}+ b_2
H^{-\frac{2}{7}})(g_{55})^{-\frac{1}{2}} [\d(y)-\d(\pi-y)] + {\it Reg}\ .
\label{5Dsingular3}
\eea
The fact that four out of the five
diagonal components of the Einstein tensor contain singular terms
suggests that we might try to couple two 3-brane actions to the
bulk theory, since the singular pieces correspond to a sum of two
terms proportional to $b_1$ and $b_2$ respectively. Let us add
source actions of the type
\bea
S^{3-brane}_{5} &=& -T\int_{M_{4}}d^{4}\s
\Big[\frac{1}{2}\sqrt{-\g}
\g^{\m \n}\pt_{\m}X^{M}\pt_{\n}X^{N}g_{MN}(X)f(\f(X)) -\sqrt{-\g} \nn \\
&&\phantom{-T\int_{M_{4}}d^{4}\s\Big[} +\frac{1}{4!} \e^{\m \n \r
\t} \pt_{\m}X^{M} \pt_{\n}X^{N} \pt_{\r}X^{P} \pt_{\t}X^{Q}
A_{MNPQ}(X)\Big]\ . \eea Here $T$ denotes the tension,
$\s^{\m}$ denote the worldvolume coordinates,\footnote{With a
slight abuse of notation we choose $\m, \n, ...$ indices to denote
worldvolume directions $0,1,2,3$ in anticipation of the fact that
we will choose the static gauge later on where the coordinates of
the brane are aligned with the coordinates of the bulk.} the
$X^{M}(\s)$ functions represent the embedding of the brane in the
ambient spacetime, $\g_{\m \n}(\s)$ is the worldvolume metric on
the brane and we have allowed for an as yet unspecified coupling
to the scalar field via $f(\f(X)).$ Also, $A_{(4)}(X)$ is a
$4$-form field that is required for consistency in a
$\mathbb{Z}_2$ symmetric background: it represents the charge of
the brane and is needed in order for the equation of motion
resulting from varying $X$ to be satisfied.

We also need to have kinetic terms for the $A_{(4)}$ fields. At this point, it is useful to remember
that  dimensional reduction of the action for the $F_{[5]}$ 5-form field strength in $D=10$ type IIB
supergravity gives rise to just such a kinetic term in $D=5$
\cite{BDLPS}. Taken by itself, this field strength describes no $D=5$ continuous degrees of freedom,
as a cursory review of its gauge structure reveals. It is a ``theory-of-almost-nothing'' field. The
caveat impled by ``almost'' is that $F_{[5]}$ couples to the 5-sphere volume modulus $\phi$, and the
integration constant arising from the $F_{[5]}$ field equation gives rise to a cosmological potential
for $\phi$.

Actually, in the $S^5$ dimensional reduction of type IIB theory,
there are two independent types of potential terms that arise for
$\phi$ -- one with a coefficient depending on the $F_{[5]}$ form
field expectation value $m$, and the other with a coefficient
determined by the value of the Ricci scalar on the 5-sphere $R_5$.
There are correspondingly two distinct types of singularity
structure that occur in the 3-brane solutions to this $D=5$
theory, as we have seen in (\ref{5Dsingular1}-\ref{5Dsingular3}).
One of these arises from the $S^5$ dimensional reduction of the
classic D3 brane of $D=10$ type IIB supergravity, while the other
can be viewed as arising from a $\mathbb{Z}_2$ identification of
two regions of $D=10$ flat space \cite{CDLLPS}. It is convenient
to introduce for this purpose a second 5-form
``theory-of-almost-nothing'' field strength in $D=5$ in order to
input the second potential coefficient from a separate brane
source coupling in a fashion similar to the way the D3 brane
charge is put in. In fact, the left and right locations of these
two brane sources can be separated. Taking $i={1,2}$ to denote the
(left, right) locations, we take the first type of sources to be
located at $X_i^M$ and the second type to be at $\tilde X_i^M$.
Both of these source actions are consistent with the general
scheme for handling supersymmetric solutions in singular spaces of
Ref.\ \cite{BKV}. We take the dimensionally reduced $F_{[5]}$ of
type IIB theory to be $F_{[5]}=dA_{[4]}$ and the second 5-form
field strength to be $\tilde F_{[5]}=d\tilde A_{[4]}$.

Then using the Israel junction conditions to determine the
coupling to the brane actions (see Appendix B), we find the
brane + bulk action 
\bea 
S_5 &=& \int d^5x \sqrt{-g}\,[R -
\frac{1}{2}(\pt \phi)^2 - \frac{1}{2\cdot 5!}e^{-8 \alpha
\phi}F_{[5]}^2 + \frac{1}{4 \cdot 4!} e^{-\frac{16}{5}\alpha
\phi}\tilde F_{[5]}^2] \nn
\\
 &&  - 4m \sum_{i=1}^2 s_i\int d^5x
\int d^4 \s \delta^5(x-X_{(i)})
[\sqrt{-\g} \g^{\mu \nu}\pt_{\mu} X_i^M \pt_{\nu} X_i^N g_{MN} e^{2
\alpha \phi} - 2\sqrt{-\g}
 \nn \\ && \qquad+ \frac{2}{4!} \e^{\m \n \r \s}\pt_{\m} X^M \pt_{\n} X^N \pt_{\r} X^P \pt_{\s} X^Q
A_{MNPQ}] \nn \\ &&  + \sqrt{5R_5} \sum_{i=1}^2 s_i\int d^5x \int
d^4 \s \delta^5(x-\tilde X) [\sqrt{-\g} \g^{\mu \nu}\pt_{\mu}
\tilde X^M \pt_{\nu} \tilde X^N g_{MN} e^{\frac{4}{5} \alpha \phi}
- 2\sqrt{-\g} \nn \\ && \qquad+ \frac{2}{4!} \e^{\m \n \r
\s}\pt_{\m} \tilde X^M \pt_{\n} \tilde X^N \pt_{\r} \tilde X^P
\pt_{\s} \tilde X^Q \tilde A_{MNPQ}] , \label{action} 
\eea 
where $s_1=1$, $s_2=-1$ give the opposing charges of the (left, right)
branes of each type. In the following, we shall take the two brane
types on each side of the interval to be coincident, \ie
$X_i^M=\tilde X_i^M$, as we are not interested here in discussing
separately the dynamics of each type. However, the general action
(\ref{action}) will be important for us as it will allow us to
discuss carefully the nature of the brane-bulk interaction and
energy conservation. For completeness we give the equations of
motion resulting from this action in Appendix A.

We may use the brane worldvolume reparameterization freedoms and $D=5$ general coordinate invariance to
choose a static gauge\footnote{As discussed in \cite{Gregory}, the static gauge for a two-brane system
does not overfix the coordinate and reparameterization gauge freedoms.} where $X_i^\mu=\sigma^\mu$,
$X_1^5=0$ and
$X_2^5=\pi$. Note that we are not fixing the physical positions of the left and right branes, just the
choice of coordinates by which we designate the two branes. Their physical location within the $D=5$
spacetime can fluctuate as a function of the bulk supergravity fields.

In the static gauge,the equations of motion for the $5$-form field strengths reduce to
\bea
\nabla_y (e^{-8 \a \phi}F_{[5]}^{y \m \n \r \s}) &=& 8m
[\delta(y)-\delta(y- \pi)]\frac{1}{\sqrt{-g}}\e^{\m \n \r \s}, \\
\nabla_y (- \frac{5}{2}e^{-\frac{16}{5} \a \phi}\tilde F_{[5]}^{y \m \n
\r \s}) &=& 2 \sqrt{5R_5} [\delta(y)-\delta(y-
\pi)]\frac{1}{\sqrt{-g}}\e^{\m \n \r \s}\ .
\eea
They have the solutions
\bea
F_{MNPQT} &=& 4m e^{8 \a \phi} \theta(y)\sqrt{-g}\,\e_{MNPQT} \label{Fsol}\\
\tilde F_{MNPQT} &=& - \frac{2}{5} \sqrt{5R_5} e^{\frac{16}{5} \a \phi}
\theta(y)\sqrt{-g}\,\e_{MNPQT}\ ,\label{tFsol}
\eea
where
\bea \theta(y) &=&
\begin{cases} +1 & \mbox{for } 0 \le y < \pi \\ -1 & \mbox{for } -
\pi \le y < 0
\end{cases}
\eea and we impose the upstairs-picture identification $y \sim
y+2\pi.$ Note that this sign flipping for the $5$-forms across the
location of the brane as given in (\ref{Fsol},\ref{tFsol}) is in
agreement with what we expect for supersymmetry in a
$\mathbb{Z}_2$ symmetric singular spacetime \cite{BKV}. We will
make the correspondence to the BKVP formalism explicit in Appendix
C. We shall mainly work in static gauge in the following.

\section{Energy}

As explained in the introduction, the bulk fields effectively
contain all the necessary knowledge about the boundary branes due
to the Israel junction conditions, which are boundary conditions
as a result of the $\mathbb{Z}_2$ symmetry (see Appendix B). This
means that if we can show positivity of the energy in the bulk, we
will have fully shown the stability of this class of spacetimes.
Let us therefore proceed by first defining the energy in the bulk.

It is well known that energy is not defined in a very obvious way
in any theory containing gravity \cite{W,AD,DS}. In particular,
diffeomorphism invariance implies that the theory is invariant
under arbitrary time reparameterizations. Energy is defined for spacetimes that
admit asymptotically a timelike Killing vector field. If we want an expression for
the energy in terms of fluctuations about a background possessing a global
timelike Killing vector, we need to expand the Einstein equations
in terms of fluctuations about the background, and then separate out the
terms of quadratic and higher orders in the fluctuations from
terms that are linear in the fluctuations. The background is taken to
satisfy the field equations exactly. Let us write \bea
g_{MN}&=&g^{(0)}_{MN}+h_{MN} \\
\phi &=& \phi^{(0)}+ \phi^{(1)} \eea and similarly for all other
fields. Here $h_{MN}=g^{(1)}_{MN}$ is the fluctuation of the
metric and the superscript denotes the order in fluctuations. So
we rewrite the Einstein equations as \bea \tau_{MN} &\equiv&
T^{(2+{\rm higher})}_{MN}-G^{(2+{\rm higher})}_{MN}
 \\  &=& R^{(1)}_{MN}- \frac{1}{2}g^{(0)}_{MN}g^{(0)RS}R^{(1)}_{RS}
+\frac{1}{2}g^{(0)}_{MN}h^{RS}R^{(0)}_{RS}
-\frac{1}{2}h_{MN}g^{(0)RS}R^{(0)}_{RS}-T^{(1)}_{MN}  \\ \nn
\label{expandEinst} \eea where $\tau_{MN}$ is the energy-momentum
pseudotensor containing the contributions due to gravitational
energy.

Now $\tau_{MN}$ satisfies 
\be 
\nabla_N^{(0)} \tau^{MN} \propto
\phi^{(0),M}\times [{\rm linearised} \ \phi \ {\rm field \ equation}] 
\ee
which is non-zero in general since we cannot impose the linearised
matter field equations (this would be in conflict with the fact
that we are imposing the full Einstein and matter field
equations). However, since our background satisfies \be
\phi^{(0),M} \xi^{(0)}_{M} = 0, \ee where $\xi_M^{(0)}$ is a
timelike background Killing vector, we have it that\footnote{A
similar line of reasoning was advocated by Deser and Soldate in
their discussion of the energy of the Kaluza-Klein monopole
\cite{DS}.} \be (\nabla_N^{(0)} \tau^{MN}) \xi_M^{(0)} = 0. \ee
Using the defining property of the Killing vector \be
\nabla^{(0)}_M \xi^{(0)}_N + \nabla^{(0)}_N \xi^{(0)}_M = 0, \ee
we can construct the ordinarily conserved vector density \cite{AD}
\be \nabla_N^{(0)}(\sqrt{-g^{(0)}}\tau^{MN}\xi^{(0)}_M) =
\pt_N(\sqrt{-g^{(0)}}\tau^{MN}\xi^{(0)}_M) =0\ . \ee This then
enables us to define the energy as \be Q= \int_V dV
{\sqrt{-g^{(0)}} \tau^{0M}\xi^{(0)}_{M}}\ , \label{EnergyInt} \ee
where $dV$ is a 4-spatial volume element. We note that for the
solution (\ref{Solution}) we have $\xi^{(0)M}=(1,\underline{0})$.
We can then look at the conservation of the energy by calculating
(where $i,j,k,...$ indices denote spatial directions) \bea
\frac{\pt}{\pt t}Q &=& - \int dV
\pt_i[\sqrt{-g^{(0)}}(G^{(1)Mi}-T^{(1)Mi})\xi^{(0)}_M] \\ &=& -
[\sqrt{-g^{(0)}}(G^{(1)05}-T^{(1)05})\xi_0^{(0)}]_{y=0}^{y=\pi} \\
&=& 0 \eea The last line follows because $G^{(1)05}$ and
$T^{(1)05}$ are continuous and odd under the $\mathbb{Z}_2$
symmetry and thus vanish at the location of the branes. The fact
that they are continuous can be explained by observing that the
brane energy-momentum tensor is given by \be T^{05}_{brane}
\propto \g^{\m \n}\pt_{\m}X^0 \pt_{\n}X^5, \ee as can be read off
from (\ref{Einstein}) in Appendix A. Thus in the static gauge
$X^{\m}= \s^{\m}, \, X^5 = 0,\pi$ we have \be T^{05}_{brane} =0.
\ee This holds at every order in perturbation theory, and it shows
that there are no singular contributions to the $05$ Einstein
equation. Moreover this shows that the bulk energy is conserved without
any contribution from the brane variables.

Before proceeding, let us first give here the ADM surface form of the energy \cite{DT}. Indeed,
$\tau^{0M}\xi^{(0)}_M$ can be rewritten as a total derivative, thus yielding the surface form (where we
have dropped the $^{(0)}$ superscript on background fields in order to avoid cluttering the expression):
\bea
Q&=&\frac{1}{2} \int_{\pt V} d
\Sigma_{i}(\xi_{N}h^{iN;0}-\xi_{N}h^{0N;i}+\xi^0 h^{,i} -
\xi^{i}h^{,0}+h^{0N}{\xi_{N;}}^{i} -h^{iN}{\xi_{N;}}^{0} \nn\\
&&\phantom{\frac{1}{2} \int_{\pt V} d
\Sigma_{i}(\xi_{N}} + \xi^{i} {{h}^{0N}}_{;N}-\xi^{0}h^{iN}_{~~;N}+ h
\xi^{i;0}+\xi^{0}\f^{,i}\f^{(1)} -\xi^{i}\f^{,0}\f^{(1)})
\label{SurfaceEnergy}
\eea
where the semicolons denote covariant differentiation with respect to the background
metric. In a similar way, one can define three total momentum charges associated
with the spacelike Killing vectors corresponding to the spatial worldvolume
translational symmetries of the background.

\section{Positive Energy}

As we saw in the last section, we can look at the bulk alone in
order to prove the stability of the $5$-dimensional HW spacetimes.
Thus, if the energy can be shown to be positive at a
given time, it will remain so due to the bulk field equations
alone, with no contribution from the boundary $X^{\mu}$ variables.
In the bulk, we can rewrite our theory (\ref{Lagrangian}) in terms
of a potential derived from a superpotential
\bea
\mathcal{L}&=&\sqrt{-g}\,[R-\frac{1}{2}(\pt
\phi)^{2}-V(\phi)] \label{gravscalar}\\
W(y,\phi) &=& \sqrt{2}(2m e^{4 \alpha \phi} - 5\sqrt{\frac{R_5}{20}}e^{\frac{8}{5}\alpha \phi})
\theta(y)
\eea
where
\bea
V(\phi) &=& W_{, \phi}^2 - \frac{2}{3} W^2
\\ &=& 8m^2 e^{8 \alpha \phi}- R_5 e^{\frac{16}{5}\alpha \phi}\ .
\eea
When extended to include fermions, the theory is invariant under the supersymmetry transformations
\cite{LiuSati}
\bea
\delta \psi_M &\equiv& \mathcal{D}_M \e = [\nabla_M - \frac{1}{6 \sqrt{2}}\G_M W(y,\phi)]\e
+ \hbox{higher order in fermions} \label{gravitino}\\
\delta \l &=& (\frac{1}{2}\G^M \nabla_M \phi + \frac{1}{\sqrt{2}}
W_{,\phi})\e + \hbox{higher order in fermions}\ . \label{dilatino}
\eea
We can then prove positivity of the energy of a purely bosonic solution to the theory
(\ref{gravscalar}) at all orders using a Witten-Nester type
argument as developed in several stages in Refs
\cite{Deser:1977hu,Witten,Nester,Deser:1983rn,Boucher,Townsend,FNSS}. Let us
define the Witten-Nester energy (we will show later on that this definition is
equivalent to the one given in the previous section) by
\be
E_{{\rm WN}} = \int_{\pt V}
* E \label{WNenergy}
\ee
where the integral is taken over the boundary of the spatial volume element $V$,
and where $*E$ is the Hodge dual of the Nester 2-form $ E =
\frac{1}{2} E_{MN} dx^M dx^N$, defined by
\be
E^{MN} = \bar\eta \,
\Gamma^{MNP} \mathcal{D}_P \eta - \overline{\mathcal{D}_P \eta} \,
\Gamma^{MNP} \eta
\ee
where $\eta$ denotes here a commuting spinor function that asymptotically tends to a background Killing
spinor, \ie it satisfies
\bea
\mathcal{D}^{(0)}_M
\eta &=& 0 \label{KillSpin1}
\\ \frac{1}{2}\G^M \nabla^{(0)}_M \phi^{(0)}\eta + \frac{1}{\sqrt{2}}
W_{,\phi}^{(0)}\eta &=& 0 \label{KillSpin2}
\eea
asymptotically as $|x^{1,2,3}|\rightarrow\infty$. The anticommuting supersymmetry parameter appearing in
the fermion transformations (\ref{gravitino},\ref{dilatino}) is given by $\eta$ times an anticommuting
constant.

We can next use Gauss' Law to rewrite the energy as an integral
over $V$ (where $d\Sigma_0 = dV$) 
\bea 
E_{{\rm WN}} &=& \int_{V}
d\Sigma_M \sqrt{-g}\,\nabla_N E^{MN} \\ &=& \int_V d\Sigma_M
\sqrt{-g}\,[\overline{\mathcal{D}_N \eta }\G^{MNP}\mathcal{D}_P \eta
+ \bar{\eta}\G^{MNP}\mathcal{D}_N\mathcal{D}_P \eta + {\rm h.c.}]
\eea 
The second term can be rewritten in terms of the dilatino
supersymmetry transformation (\ref{dilatino}). If we now impose
the Witten condition \be \Gamma^{k} \mathcal{D}_{k} \eta =0\
,\label{wittencond} \ee and choose to foliate our spacetime in
terms of spatial slices at constant times, we can express the
Witten-Nester energy in terms of $\tilde\delta\psi_i$,
$\tilde\delta\lambda$, which are related to $\delta\psi_i$ and
$\delta\lambda$ in (\ref{gravitino},\ref{dilatino}) by replacing
the anticommuting spinor parameter $\epsilon$ by the commuting
$\eta$, which is subject to the Witten condition
(\ref{wittencond}): \be E_{{\rm WN}} = 2 \int_V dV
\sqrt{-g}\,[(\tilde\delta \psi_i)^{\dag}  \tilde\delta \psi_i +
(\tilde\delta \l)^{\dag} \tilde\delta \l] \ge 0\ . \ee

What remains to be done is to show that the energy definition
given in (\ref{WNenergy}) actually agrees with the previous
definition (\ref{EnergyInt}). Let us expand the expression
(\ref{WNenergy}) in fluctuations about the background by
perturbing the vielbeins ${e^P}_a= {e^{(0)P}}_a +
\frac{1}{2}{h^P}_a,$ where $a,b,...$ denote tangent space indices.
For the spin connection we find \be \omega^{(1)}_{Pab}=
\frac{1}{2}(h_{Pa;b}-h_{Pb;a}) \ee and we also expand \be
\mathcal{D}_P \eta= \frac{1}{4} \omega^{(1)}_{Pab}\G^{ab}\eta-
\frac{1}{6 \sqrt{2}}\G_P W_{,\phi}^{(0)} \phi^{(1)}\eta-
\frac{1}{12 \sqrt{2}}h_{Pa} \G^a W^{(0)}\eta + ... \ee for a
Killing spinor $\eta.$ The energy (\ref{WNenergy}) can then be
written as \bea E_{{\rm WN}} = \int_{\pt V}&&[\frac{1}{4} \bar\eta
\Gamma^M \eta(h^{;N}-{h_P}^{N;P}) - \frac{1}{4}\bar\eta\Gamma^N
\eta (h^{;M}-{h_P}^{M;P}) \nn \\ &&+\frac{1}{4}\bar\eta\Gamma^P
\eta ({h_P}^{N;M}-{h_P}^{M;N}) \nn \\ && - \frac{1}{12
\sqrt{2}}\bar\eta(\G^{MN}h+\G^{NP}{h_M}^P+\G^{PM}{h_P}^N)W(\phi)\eta
\nn \\ &&- \frac{1}{2 \sqrt{2}}\bar\eta\G^{MN} W_{, \phi}
\phi^{(1)} \eta] d\Sigma_{MN} +h.c. \eea We now relate Killing
spinors to Killing vectors by \be \xi^{(0)M} = \bar\eta \G^M \eta.
\ee Using relations (\ref{KillSpin1},\ref{KillSpin2}) one can
see that the Witten-Nester surface expression for the energy is
exactly equivalent to the ADM type formula (\ref{SurfaceEnergy}),
with the terms of the form $\bar\eta \G^{MN} W \eta$ proportional
to terms involving $\xi^{(0)N;M},$ and the terms of the form
$\bar\eta \G^{MN} W_{,\phi}\eta$ responsible for the
$\xi^{(0)[M}\phi^{(0),N]}$ contributions. Thus we have verified
that our two definitions of the energy are in agreement. We can
therefore conclude that the energy is manifestly positive and that
the HW spacetimes are stable despite the presence of brane sources
of negative tension\footnote{We note
that the analogous four dimensional Schwarzschild
solution with negative mass parameter has recently been shown
to be stable subject to linearised perturbations of finite total
energy in Ref.\ \cite{Gibbons:2004au}}.

\section{Positivity at Quadratic Order}

The Witten-Nester argument just presented is very powerful, albeit rather
non-intuitive. We feel that it is sometimes good to also have a more concrete,
although less general, argument and therefore we wish to show here, by means of
explicit illustration, that the energy is positive at quadratic order in
perturbation theory \cite{Brill}. In fact, any potential instability would
presumably manifest itself already at this order, and therefore
stability at quadratic order can already be seen as a strong indication
of stability to all orders. For our calculations in this
section we will use the ADM Hamiltonian approach \cite{ADM}, as it
is best suited for this purpose. We perform an
explicit $(1+4)$ decomposition of the metric, choosing spacetime
to be foliated along constant time slices. The metric can be written as
\be
ds^{2}=(N_{i}N^{i}-N^{2})dt^2+2N_i dx^i dt
+g_{ij}dx^i dx^j\ ,
\ee
where $N$ is the lapse function, $N^i$ the
shift function and $g_{ij}= {}^{(5D)}g_{ij}$ (see \cite{MTW} for
details). Indices are lowered and raised by $g_{ij}$ and its
inverse $g^{ij}$ (which does not equal $^{(5D)}g^{ij}$ in
general!). A dot on top of a quantity denotes a time derivative,
while $_{\mid}$ denotes covariant differentiation with respect to
the 4-dimensional metric. The embedding of the
4-dimensional hypersurface in the 5-dimensional bulk spacetime is characterised by
the extrinsic curvature $K_{ij}$, given as 
\be
K_{ij}=\frac{1}{2N}(-\dot{g_{ij}}+N_{i\mid j}+N_{j\mid i})\ .
\ee
The ``momentum''conjugate to the metric is defined as \be
\pi^{ij}\equiv \frac{\delta \mathcal{L}}{\delta
\dot{g_{ij}}}=-g^{\frac{1}{2}}(K^{ij}-g^{ij}K)\ ,
\ee 
and the momentum $P$ conjugate to the scalar field $\f$ reads 
\be 
P\equiv
\frac{\delta\mathcal{L}}{\delta\dot{\f}}=\frac{g^{\frac{1}{2}}}{N}(\dot{\f}-N^i
\f_{\mid i})\ . 
\ee In terms of these new variables, the action
can be rewritten \cite{JJH}
\bea
S&=&\int dt d^{4}x
\{\pi^{ij}\dot{g_{ij}}+P \dot{\f} \nn \\
&& \phantom{\int dt d^{4}x}-N g^{-\frac{1}{2}}[
\pi^{ij}\pi_{ij}-\frac{1}{3}\pi^2-g(^{(4D)}R- V(\f))
+\frac{1}{2}P^2+\frac{1}{2}g g^{ij}\f_{\mid i}\f_{\mid j}]
 \\ \nn
&& \phantom{\int dt d^{4}x}-N_{i}[-2{\pi^{ij}}_{\mid j}+\f^{\mid i} P]\}
\eea
We see that $N$ and $N_i$ act as Lagrange multipliers,
imposing respectively the constraints
\bea
&&
\pi^{ij}\pi_{ij}-\frac{1}{3}\pi^2-g(^{(4D)}R- V(\f))
+\frac{1}{2}P^2+\frac{1}{2}g g^{ij}\f_{\mid i}\f_{\mid j}=0  \\
&& -2{\pi^{ij}}_{\mid j}+\f^{\mid i} P=0\ .
\eea
At background order, these constraints are of course satisfied by the
solution (\ref{Solution}), with
\be
^{(4D)}R=\frac{1}{2}g^{(0)ij}\f^{(0)}_{\mid i}\f^{(0)}_{\mid
j}+V(\f)
\ee
and
\be
0=P^{(0)}=\pi^{(0)ij}=N^{(0)i}=\dot{\f}^{(0)}\ .
\ee
We impose
these constraints at linear order, where they read
\bea
&&
-^{(4D)}R^{(1)}+\frac{\pt V}{\pt
\f}\f^{(1)}-\frac{1}{2}h^{ij}\f^{(0)}_{\mid i}\f^{(0)}_{\mid
j}+g^{(0)ij}\f^{(0)}_{\mid i}\f^{(1)}_{\mid j}=0   \\
&& 2{\pi^{(1)ij}}_{\mid j}=\f^{(0)\mid i} P^{(1)}\ ,
\eea
with \be ^{(4D)}R^{(1)}={h^{ij}}_{\mid ji}-{{h^{i}}_{i\mid
j}}^{j}-h^{ij(4D)}R^{(0)}_{ij}. \ee \\
If we write the action in the form 
\be 
S=\int dt d^{4}x\,
[\pi^{ij}\dot{g_{ij}}+P\dot{\f}-NH-N_i H^i] , 
\ee 
we can read
off the Hamiltonian 
\be 
\mathcal{H}=\int d^{4}x\,[NH+N_iH^i]\ .
\ee

To second order, subject to the constraints at linear order, we
then find
\bea
\mathcal{H}^{(2)} &=& \int d^{4}x \{
N^{(0)}g^{(0)-\frac{1}{2}}[\pi^{(1)ij}\pi^{(1)}_{ij}-\frac{1}{3}\pi^{(1)2}+\frac{1}{2}P^{(1)2}]
\nn \\ && + N^{(0)}g^{(0)\frac{1}{2}}[\frac{1}{4}h^{ij\mid
k}h_{ij\mid k} +\frac{1}{4}{{h^{i}}_{i\mid}}^{j}{h^{k}}_{k\mid j}
-\frac{1}{2}{h^{ij}}_{\mid j}{h_{ik\mid}}^{k}-\frac{1}{12}h^{ij}h_{ij}(V+\frac{1}{2}g^{(0)kl}\f^{(0)}_{,k}\f^{(0)}_{,l})] \nn \\
&& +N^{(0)}g^{(0) \frac{1}{2}}[\frac{1}{2} \frac{\pt^2 V}{\pt
\f^2}\f^{(1)2}+\frac{1}{2}(\f^{(1)\mid i}-h^{ij}\f^{(0)}_{\mid j
})(\f^{(1)}_{\mid i}-{h_{i}}^{k}\f^{(0)}_{\mid k})] \}\ . \eea It
is straightforward to verify that the second order Hamiltonian is
conserved in time subject to the linearised field equations. We
can see that its expression exhibits terms of various signs.
However, before drawing any premature conclusions, we must
remember that no gauge choice has yet been imposed, and therefore
there is still a large amount of ambiguity in the precise meaning
of the above expression. We have $5$ gauge choices at our
disposal. Let us choose: \bea && {h^{ij}}_{\mid j}=0, \\ &&
P^{(1)2} = \frac{2}{3} \pi^{(1)2}+g^{(0)}\mid \frac{\pt^2 V}{\pt
\f^2} \mid \f^{(1)2}\ . \eea These gauge choices are
non-conflicting and are independent of each other. Now note that the
requirement $V+\frac{1}{2}g^{(0)ij}\f^{(0)}_{\mid i}\f^{(0)}_{\mid
j}\leq 0$ for the background translates into \be \frac{3
k^2}{196H}(16 b_1^2 H^{-\frac{3}{7}}+20 b_1 b_2 - 5 b_2^2
H^{\frac{3}{7}})\leq 0\ , \ee and can be verified to be always
satisfied if the reality condition (\ref{Reality}) on the metric
is imposed. Therefore, our expression for the second order
Hamiltonian is manifestly positive, indicating that this system is
stable, despite the presence of the negative-tension domain wall.

\section{Discussion}
We have considered the energy and the stability of
Ho\v{r}ava-Witten spacetimes, and we have shown that it is
essentially the unbroken supersymmetry of the theory and of the static background solution
that guarantees their stability: the supersymmetric background acts like a ``vacuum'' whose energy bounds
from below that of neighbouring perturbations. This happens despite the presence of negative tension branes
which might have been thought to give rise to unstable modes. In fact, the dynamics of the branes can be
studied {\it via} the dynamics of the bulk fields at and near the
location of the boundaries. This is possible because the world volume {\it
local} $\mathbb{Z}_2$ symmetry of the solutions that we consider gives
rise to boundary conditions relating brane and bulk.

In fact, one might wonder what would happen if one were to relax
altogether the requirement of a local $\mathbb{Z}_2$ symmetry.
Perturbations of this type would change the topology of the
solutions considered to $S^1 \times \mathbb{R}^4$, and such
perturbations are not included in the analysis of this paper. In
that case, the matching conditions at the branes would not simply
reduce to boundary conditions and our expression for the bulk
energy would not be a separately conserved quantity anymore. The
threat of unstable modes from the negative tension brane is thus
likely to become much more substantial in that case.

\section*{Acknowledgments}
We would like to thank Paul Bostock, Thibault Damour, Stanley
Deser, Gilles Esposito-Farse, Ruth Gregory, Jonathan Halliwell,
Chris Hull, Jussi Kalkkinen and Chris Pope for clarifying
discussions. KSS would like to thank the IHES in Bures-sur-Yvette
and the Albert-Einstein Institute in Potsdam for hospitality. We
all would like to thank CERN for hospitality; and we would like to
acknowledge the support of PPARC. JLL acknowledges the support of
a Bourse Formation-Recherche of the Luxembourgish Minist\`{e}re de
la Culture, de l'Enseignement Sup\'{e}rieur et de la Recherche,
and of a University of London Valerie Myerscough Prize. PS is
supported by the Leverhulme Trust.

\section*{Appendix A}

We denote spacetime coordinates by
$(x^0,x^1,x^2,x^3,y)=(0,1,2,3,5).$ Indices $M,N,P,...$ run over
all coordinate values; $a,b,...$ denote flat tangent space
indices, whereas $\mu, \nu, ...$ run over $0,1,2,3$ and $i,j,...$
are purely spatial and take the values $1,2,3,5.$ Our metric has
signature $(-,+,+,+,+)$ and our conventions for gravity are
\bea
&& R_{MN} = \pt_P \G^P_{MN} - \pt_M \G^P_{NP} + \G^P_{MN}
\G^L_{PL} - \G^L_{MP} \G^P_{NL} \\ && [\nabla_M ,\nabla_N] V_P =
{R_{MNP}}^L V_L
\eea

Let us also write down the perturbed expressions
\bea
\G^{(1)P}_{MN} &=& \frac{1}{2}({h^P}_{M;N}+{h^P}_{N;M}-{h_{MN;}}^P) \\
R^{(1)}_{MN} &=& \frac{1}{2}({h^P}_{M;NP}+{h^P}_{N;MP}
-{h^P}_{P;MN}-{h_{MN;P}}^P) \eea For a Killing vector $\xi^M$, by
definition we have $\xi_{M;N}+\xi_{N;M}=0,$ and from this we can
derive the useful identity \be \xi_{M;NP}={R_{MNP}}^L \xi_L\ . \ee
Another useful relation in deriving the surface expression
(\ref{SurfaceEnergy}) is \be \xi^{R;M}\phi^{(0)}_{,R}+\xi^R
{\phi^{(0);M}}_{R} = \nabla^M(\xi^R \phi^{(0)}_{,R}) = 0 = \xi^R
\phi^{(0)}_{,R} \ee The $D=5$ Gamma matrices satisfy \be \G^{MNP}
\G_{T} = {\G^{MNP}}_T + 3!\G^{[MN} {\d^{P]}}_T, \ee and repeated use
of this identity leads to the useful formula \be \G^{MNP} \G^{TS}
R_{NPTS} = 4 \G^N {G_N}^M\ . \ee Let us also note the basic
identity \be [\nabla_M,\nabla_N] = \frac{1}{4}R_{MNAB}\G^{AB} \ee
and the useful expression \be \G^N \G^M \G^P \phi^{(0)}_{,N}
\phi^{(0)}_{,P} = 2 \phi^{(0),M}\G^P \phi^{(0)}_{,P} - \G^M
\phi^{(0),P} \phi^{(0)}_{,P}. \ee For an action of the type 
\bea
S_5 &=& \int d^5x \sqrt{-g}\,[R - \frac{1}{2}(\pt \phi)^2 -
\frac{c}{2\cdot 5!}e^{a
\phi}F_{(5)}^{2}]  \nn \\
 && - T \int d^5x \int d^4 \s \delta^5(x-X)
\big[\sqrt{-\g} \g^{\mu \nu}\pt_{\mu} X^M \pt_{\nu} X^N g_{MN}
f(\phi(x))- 2\sqrt{-\g}
 \nn \\ && \phantom{- T \int d^5x \int d^4 \s \delta^5(x-X)} + \frac{2}{4!} \e^{\m \n \r \s}\pt_{\m} X^M
\pt_{\n} X^N \pt_{\r} X^P \pt_{\s} X^Q A_{MNPQ}(x)\big]
\label{apaction}
\eea
where for simplicity we consider here just a single brane source,
we have the equations of motion
\bea
G^{MN} &=& \frac{1}{2}\phi^{,M}\phi^{,N} - \frac{1}{4}g^{MN}
\phi^{,P}\phi_{,P} + \frac{c}{2 \cdot 4!}e^{a \phi}(F^2)^{MN} -
\frac{c}{4 \cdot 5!} e^{a \phi} F^2 g^{MN} \nn \\ && - T
\frac{1}{\sqrt{-g}} \int d^4 \s \delta^5(x-X) \sqrt{-\g} \g^{\mu
\nu}\pt_{\mu} X^M \pt_{\nu} X^N f(\phi) \label{Einstein} \\
\Box \phi &=& \frac{ac}{2 \cdot 5!} e^{a \phi} F^2 \nn \\ &&+ T
\frac{1}{\sqrt{-g}} \int d^4 \s \delta^5(x-X) \sqrt{-\g} \g^{\mu
\nu}\pt_{\mu} X^M
\pt_{\nu} X^N g_{MN} \frac{\pt f}{\pt \phi} \\
\nabla_M(e^{a \phi}F^{MNPQR})&=& \frac{2T}{c\sqrt{-g}} \int d^4 \s
\d^5(x-X) \e^{\m \n \r
\t}\pt_{\m}X^N\pt_{\n}X^P\pt_{\r}X^Q\pt_{\t}X^R \\ \g_{\m \n} &=&
\pt_{\m}X^M \pt_{\n}X^N g_{MN} f(\phi) \\
0 &=& \pt_{\m} (\sqrt{-\g}\g^{\m \n}\pt_{\n}X^N f(\phi)) +
\sqrt{-\g}\g^{\m \n}\pt_{\m}X^P\pt_{\n}X^Q \G^N_{PQ}f(\phi) \nn \\
&& -\frac{1}{2}\sqrt{-\g}\g^{\m \n}\pt_{\m}X^P\pt_{\n}X^Q g_{PQ}
\frac{\pt f}{\pt \phi} \phi^{,N} \nn \\ && - \frac{1}{4!} \e^{\m
\n \r \t}\pt_{\m}X^P\pt_{\n}X^Q\pt_{\r}X^R\pt_{\t}X^S {F^N}_{PQRS}
\eea

\section*{Appendix B}

For simplicity we will present the junction conditions here for
the theory specified by the action (\ref{apaction}), and we will
work in the static gauge $X^{\m}= \s^{\m}, \, X^5 = 0,\pi$. Then
there are no junction conditions associated with the $55$ and $\mu
5$ components of the Einstein equations, since \be T^{55}_{brane}
= 0 = T^{\mu 5}_{brane}. \ee However $T^{\mu \nu}_{brane}$ is
non-zero and singular, and we can derive the associated junction
conditions by integrating the $\mu \nu$ components of the Einstein
equations across the brane hypersurface. From 
\be 
G_{\mu \nu} = -
\frac{T}{\sqrt{g_{55}}}\delta(y)g_{\mu \nu}f^2(\phi) + {\it Reg}
\ee 
we have 
\be 
R_{\mu \nu} = \frac{T}{3\sqrt{g_{55}}}\delta(y)
g_{\mu \nu}f^2(\phi) + {\it Reg}\ ; 
\ee 
then we integrate over the
only component of the Ricci tensor that is singular, \ie 
the component with two $y$ derivatives. This is the only component
that can give a singular contribution if the metric is continuous:
\be \int_{-\epsilon}^{+\epsilon}dy \ (-\frac{1}{2} g_{\mu \nu,yy})
= \int_{-\epsilon}^{+\epsilon}dy \ \frac{T}{3}\sqrt{g_{55}}\,g_{\mu
\nu} f^2(\phi)\delta(y), \ee after which we take the limit
$\epsilon \rightarrow 0.$ Usually this would give us an expression
for the jump in the normal derivative of the metric across the
brane, but because of the $\mathbb{Z}_2$ symmetry that we are
imposing about the brane, the normal derivative of the metric
takes opposite values on opposite sides of the brane, and thus we
find that not just its difference, but actually its total value is
related to the value of the fields on the brane: 
\be 
g_{\mu\nu,y}\big{|}_{y=0} = - \frac{T}{3}\sqrt{g_{55}}\,g_{\mu\nu}
f^2(\phi)\big{|}_{y=0}\ . 
\ee 
A similar junction conditions can also
be derived for the scalar field: 
\be 
\phi_{,y}\big{|}_{y=0} = 2T
\sqrt{g_{55}}\,f \frac{\pt f}{\pt \phi}\big{|}_{y=0}\ . 
\ee

\section*{Appendix C}
Here we reproduce the total action for the bulk plus two brane system written in
the conventions of \cite{BKV}; this system is equivalent to (\ref{action}).
Note that the brane action is written in Nambu-Goto form:
\bea
S &=& \int d^5x \sqrt{-g}\,\left(R-\frac{1}{2}(\pt
\phi)^{2}-V(\phi,x)\right) \nonumber \\&+& \frac{1}{6} \int d^5x
~\e^{MNPQR} A_{MNPQ}\, \pt_{R}m(x)
 \nonumber\\
&-&8m\int d^5x \sum_{i=1}^2 s_i \int d^4\sigma \delta^5(x-X_i)
\left(\sqrt{-\gamma}\,\tilde W(\phi(x)) + \frac{1}{4!} \e^{\m \n \r \s} A_{\m
\n \r \s}(x) \right)\ ,
\eea
where $s_{1,2}=\pm 1$ as before and now
\bea
V(\phi,x) &=& 8m(x)^2\left(\tilde W(\phi)_{,
\phi}^2 - \frac{2}{3} W'(\phi)^2\right) \\
\tilde W(\phi(x)) &=& \frac{W}{2\sqrt2 m} = e^{4 \alpha \phi(x)} -
\frac{5}{2}\sqrt{\frac{R_5}{20~m^2}}e^{\frac{8}{5}\alpha \phi(x)}
\eea 
Written in static gauge, the important equations of motion
are: 
\bea
\partial_ym(y) &=& 2~m~\delta(y)\\
F_{MNPQR} &\equiv& 5\partial_{[M}A_{NPQR]}~=~
\sqrt{-g}\,\frac{V(x,\phi)}{2~m(x)} \e_{MNPQR}\ .
\eea

\end{document}